\RequirePackage{fix-cm}
\documentclass[twocolumn,epjc3]{svjour3}
\smartqed
\RequirePackage{graphicx}

\usepackage{amsmath,amssymb,amsfonts}
\usepackage{graphicx}
\usepackage{textcomp,nicefrac}
\usepackage{algorithmic}
\usepackage{algorithm}
\usepackage{array}
\usepackage[caption=false,font=normalsize,labelfont=sf,textfont=sf]{subfig}
\usepackage{stfloats}
\usepackage{url}
\usepackage{verbatim}
\usepackage[pdftex,draft=false, debug=false, hyperindex, colorlinks=true, linkcolor=black, citecolor=blue, urlcolor=blue, linktoc=all]{hyperref}
\usepackage{multirow}
\usepackage{orcidlink}
\usepackage{microtype}

\newcommand{\mbb}       {\ensuremath{m_{\beta\beta}}}

\newcommand{\onbb}      {\ensuremath{0\nu\beta\beta}}

\newcommand{\bb}        {\ensuremath{\beta\beta}}

\newcommand{\qbb}       {\ensuremath{Q_{\beta\beta}}}

\newcommand{\ctsper}    {cts/(keV\,kg\,yr)}

\newcommand{\tyr}       {t$\,$yr}

\newcommand{\nuc}[2]    {$^{#1}$\textrm{#2}} 
\newcommand{\gesix}     {\nuc{76}{Ge}}

\newcommand{\ppc}       {PPC}
\newcommand{\ic}       {IC}
\newcommand{\bege}      {BEGe}
\newcommand{\coax}      {Coax}
\newcommand{\pygama}      {\texttt{pygama}}

\newcommand{\LEG}       {LEGEND}
\newcommand{\Ltwo}      {{\LEG-200}}
\newcommand{\Lthou}     {{\LEG-1000}}

\newcommand{\MJD}       {\textsc{Majorana Demonstrator}}
\newcommand{\Gerda}     {\textsc{Gerda}}
 
\usepackage[numbers,sort&compress]{natbib}
\usepackage[switch,pagewise]{lineno}
\usepackage{orcidlink}
\usepackage[T1]{fontenc}
\usepackage{tikz}
\usepackage{siunitx}

\journalname{Eur. Phys. J. C}

\begin{document}

\onecolumn

\title{Energy Calibration and Performance of HPGe Detectors in the LEGEND-200 Experiment}
\titlerunning{Energy Calibration and Performance in LEGEND-200}

\thankstext{e1}{Correspondence: editorial-board@legend-exp.org}
\thankstext{e2}{Institutional Board membership suspended since April 26, 2022.}
\thankstext{e3}{Present Address: University of Delaware, Newark, Delaware 19713, USA}
\thankstext{e4}{Present Address: Center for Fundamental Physics, Northwestern University, Evanston, IL 60208, USA}
\thankstext{e5}{Present address: Air Force Institute of Technology, Dayton, Ohio 45433, USA}
\thankstext{e6}{Present address: Space Sciences Laboratory, University of California, Berkeley, CA 94720, USA}
\thankstext{e7}{Present address: Department of Physics, Simon Fraser University, Burnaby, BC 8888, Canada}


\newcommand{\MPP}{Max-Planck-Institut f\"{u}r Physik, Garching b. M\"{u}nchen, 85748, Germany}
\newcommand{\UNM}{Department of Physics and Astronomy, University of New Mexico, Albuquerque, NM 87131, USA}
\newcommand{\LAquila}{Dipartimento di Scienze Fisiche e Chimiche dell'Universit\`{a} degli Studi dell'Aquila, L'Aquila, 67100, Italy}
\newcommand{\GSSI}{Gran Sasso Science Institute, L'Aquila, 67100, Italy}
\newcommand{\LNGS}{Istituto Nazionale di Fisica Nucleare, Laboratori Nazionali del Gran Sasso, 67100 Assergi (AQ), Italy} 
\newcommand{\UTAustin}{Department of Physics, University of Texas at Austin, Austin, TX 78712, USA}
\newcommand{\LBNLNSD}{Institute for Nuclear and Particle Astrophysics and Nuclear Science Division, Lawrence Berkeley National Laboratory, Berkeley, CA 94720, USA}
\newcommand{\LBNLENG}{Engineering Division, Lawrence Berkeley National Laboratory, Berkeley, CA 94720, USA}
\newcommand{\IKZ}{Leibniz-Institut f\"{u}r Kristallz\"{u}chtung, Berlin, D-12489, Germany}
\newcommand{\IU}{Center for Exploration of Energy and Matter, and Department of Physics, Indiana University, Bloomington, IN 47405, USA}
\newcommand{\Bratislava}{Department of Nuclear Physics and Biophysics, Comenius University, Bratislava, SK-84248, Slovakia}
\newcommand{\SFU}{Department of Chemistry, Simon Fraser University, Burnaby, British Columbia, V5A 1S6, Canada}
\newcommand{\UNC}{Department of Physics and Astronomy, University of North Carolina, Chapel Hill, NC 27599, USA}
\newcommand{\Warwick}{Department of Physics, University of Warwick, Coventry, CV4 7AL, United Kingdom}
\newcommand{\TUNL}{Triangle Universities Nuclear Laboratory, Durham, NC 27708, USA}
\newcommand{\Duke}{Department of Physics, Duke University, Durham, NC 27708, USA}
\newcommand{\USC}{Department of Physics and Astronomy, University of South Carolina, Columbia, SC 29208, USA}
\newcommand{\Jag}{M. Smoluchowski Institute of Physics, Jagiellonian University, Cracow, 30-348, Poland}
\newcommand{\Dresden}{Technische Universit\"{a}t Dresden, Dresden, 01069, Germany}
\newcommand{\JINR}{Joint Institute for Nuclear Research, Dubna, 141980, Russia} 
\newcommand{\INRRAS}{Institute for Nuclear Research of the Russian Academy of Sciences, Moscow, 119991, Russia}
\newcommand{\Geel}{European Commission, Joint Research Centre (JRC), Geel, 2440 Belgium}
\newcommand{\MPIK}{Max-Planck-Institut f\"{u}r Kernphysik, Heidelberg, 69117, Germany}
\newcommand{\Queens}{Department of Physics, Engineering Physics \& Astronomy, Queen's University, Kingston, Ontario, K7L 3N6, Canada} 
\newcommand{\UTK}{Department of Physics and Astronomy, University of Tennessee, Knoxville, TN 37916, USA}
\newcommand{\Liverpool}{University of Liverpool, Liverpool, L69 3BX, United Kingdom}
\newcommand{\UCL}{Department of Physics and Astronomy, University College London, London, WC1E 6BT, United Kingdom}
\newcommand{\LANL}{Los Alamos National Laboratory, Los Alamos, NM 87545, USA}
\newcommand{\MILB}{Universit\`{a} degli Studi di Milano Bicocca, Milan, 20126, Italy}
\newcommand{\MILBINFN}{Istituto Nazionale di Fisica Nucleare, Sezione di Milano Bicocca, Milan, 20126, Italy}
\newcommand{\MILC}{Universit\`{a} degli Studi di Milano, Milan, 20133, Italy}
\newcommand{\MILCINFN}{Istituto Nazionale di Fisica Nucleare, Sezione di Milano, Milan, 20133, Italy}
\newcommand{\NRCKI}{National Research Centre ``Kurchatov Institute'', Moscow, 123098, Russia}
\newcommand{\MEPhI}{National Research Nuclear University MEPhI (Moscow Engineering Physics Institute), 115409 Moscow, Russia}
\newcommand{\TUMPhy}{Department of Physics, TUM School of Natural Sciences, Technical University of Munich, 85748 Garching b. M\"{u}nchen, Germany}
\newcommand{\ORNL}{Oak Ridge National Laboratory, Oak Ridge, TN 37830, USA}
\newcommand{\PadovaUniv}{Dipartimento di Fisica e Astronomia dell'Universit\`{a} degli Studi di Padova, Padua, 35131, Italy}
\newcommand{\PadovaINFN}{Istituto Nazionale di Fisica Nucleare, Sezione di Padova, Padua, 35131, Italy}
\newcommand{\CTU}{Czech Technical University in Prague, Institute of Experimental and Applied Physics, CZ-11000 Prague, Czech Republic}
\newcommand{\Roma}{Universit\`{a} degli Studi di Roma Tre, Rome, 00146, Italy}
\newcommand{\RomaINFN}{Istituto Nazionale di Fisica Nucleare, Sezione di Roma Tre, Rome, 00146, Italy}
\newcommand{\TTU}{Tennessee Tech University, Cookeville, TN 38505, USA}
\newcommand{\NCSU}{Department of Physics, North Carolina State University, Raleigh, NC 27607, USA}	
\newcommand{\SDSMT}{South Dakota Mines, Rapid City, SD, 57701, USA}
\newcommand{\UW}{Center for Experimental Nuclear Physics and Astrophysics, and Department of Physics, University of Washington, Seattle, WA 98195, USA}
\newcommand{\Tuebingen}{University T\"{u}bingen, T\"{u}bingen, 72076, Germany}
\newcommand{\USD}{Department of Physics, University of South Dakota, Vermillion, SD 57069, USA} 
\newcommand{\UZH}{Physik-Institut, University of Z\"{u}rich, Z\"{u}rich, 8057, Switzerland}
\newcommand{\Daresbury}{Science and Technology Facilities Council (STFC) Daresbury Laboratory, Daresbury, Cheshire, WA4 4AD, UK}
\newcommand{\Polimi}{Politecnico di Milano, Dipartimento di Elettronica, Informazione e Bioingegneria, Milan, 20133, Italy}
\newcommand{\PolimiINFN}{Istituto Nazionale di Fisica Nucleare, Sezione di Milano, Milan, 20133, Italy}
\newcommand{\LiebnitzPoly}{Leibniz-Institut f\"{u}r Polymerforschung Dresden e.V., Dresden, D-01069, Germany}
\newcommand{\Frascati}{Istituto Nazionale di Fisica Nucleare, Laboratori Nazionali di Frascati, 00044 Frascati (RM), Italy}
\newcommand{\Naples}{Dipartimento di Fisica dell'Universit\`{a} degli Studi di Napoli ``Federico II'', Naples, 80126, Italy}
\newcommand{\NaplesINFN}{Istituto Nazionale di Fisica Nucleare, Sezione di Napoli, Naples, 80126, Italy}
\newcommand{\UCSD}{Department of Physics, University of California, San Diego, La Jolla, CA 92093, USA}
\newcommand{\DataSci}{Hal\i c{\i}o\u{g}lu Data Science Institute, University of California, San Diego, La Jolla, CA 92093, USA}
\newcommand{\NTU}{Department of Physics, National Taiwan University, Taipei, 106319, Taiwan}
\newcommand{\Campania}{Dipartimento di Matematica e Fisica, Universit\`{a} degli Studi della Campania ``Luigi Vanvitelli'', Caserta, 81100, Italy}
\newcommand{\HSRM}{Faculty of Engineering, Hochschule RheinMain, Wiesbaden, 65197, Germany}
\newcommand{\PNNL}{Pacific Northwest National Laboratory, Richland, WA 99354, USA}
\newcommand{\Scuola}{Scuola Superiore Meridionale, Naples, 80134, Italy}

\author{\scriptsize
The~LEGEND~Collaboration\thanksref{e1}
\and  \\[4mm]
H.~Acharya\orcidlink{0009-0000-2219-8511}\thanksref{addr1,addr2} \and
M.~Agostini\orcidlink{0000-0003-1151-5301}\thanksref{addr3} \and
A.~Alexander\orcidlink{0000-0001-5608-1397}\thanksref{addr3} \and
C.~Alvarez-Garcia\orcidlink{0000-0003-4768-8589}\thanksref{addr4} \and
V.~Aures\thanksref{addr5,addr6} \and
F.T.~Avignone~III\orcidlink{0000-0002-5252-9104}\thanksref{addr7,addr8} \and
M.~Babicz\orcidlink{0000-0002-1017-5440}\thanksref{addr9} \and
W.~Bae\orcidlink{0000-0002-7646-7577}\thanksref{addr10} \and
M.~Balata\orcidlink{0000-0001-6745-6983}\thanksref{addr11} \and
A.S.~Barabash\orcidlink{0000-0002-5130-0922}\thanksref{addr12}
% ...
}

\institute{
\UNC \label{addr1}\and
\TUNL \label{addr2}\and
\UCL \label{addr3}\and
\MPP \label{addr4}\and
\TUMPhy \label{addr5}\and
\MPIK \label{addr6}\and
\USC \label{addr7}\and
\ORNL \label{addr8}\and
\UZH \label{addr9}\and
\UTAustin \label{addr10}\and
\LNGS \label{addr11}\and
\NRCKI \label{addr12}
% ...
}

\date{Received: date / Accepted: date}

\maketitle

\twocolumn

\begin{abstract}
This paper describes the energy scale procedures and germanium detectors performance in the \Ltwo\ experiment, a critical component for the first unblinding in the search for neutrinoless double beta decay. We detail the digital signal processing pipeline, the methodologies for peak-shape modeling and energy calibration procedures utilizing weekly $^{228}$Th source calibration runs. The optimized energy reconstruction achieves a combined average resolution of $(2.47 \pm 0.08)$~keV at $\qbb = 2039$~keV. The weekly variation of calibration peak positions are below 0.05~keV for energies up to 2614.5~keV, showing a high stability of the energy scale over time and across detectors. Furthermore, systematic corrections effectively address residual non-linearities and energy bias in the region of interest.

\end{abstract}

\section{Introduction\label{Sec:Intro}}

The dominance of matter over antimatter in our universe is one of the most interesting aspects of cosmology. One of the favored models to explain this asymmetry is leptogenesis~\cite{Fukugita:1986hr}, which is based on lepton number violation.
In many extensions of the Standard Model, neutrinos are assumed to be their own antiparticles, i.e., Majorana particles~\cite{Majorana:1937vz}, giving a possible explanation for the origin of the low neutrino mass and leading to lepton number conservation violating processes.
At present, the only feasible experiments with the potential of establishing that massive neutrinos are Majorana particles are those searching for neutrinoless double beta (\onbb) decay~\cite{Agostini:2022zub}.

The \onbb\ decay signal is predicted to appear as a narrow peak at the $Q$-value, \qbb, in the summed energy spectrum of the two emitted electrons.
In the scenario of light Majorana neutrino exchange via left-handed weak currents, the \onbb\ decay rate can be factorized into three terms:
\begin{equation}
\left(T_{1/2}^{0\nu}\right)^{-1}=G_{0\nu}|M_{0\nu}|^2 \left(\frac{m_{\beta\beta}}{m_e}\right)^2
\label{Eq:rate_0nu}
\end{equation}
where $T_{1/2}^{0\nu}$ is the half-life of the \onbb\ process, $G_{0\nu}$ is the phase space factor, $M_{0\nu}$ is the nuclear matrix element~\cite{iachello}, $m_e$ is the electron mass, and \mbb\ is the effective Majorana mass (combination of neutrino mass eigenvalues).
The key idea of these experiments is that, by studying the \onbb\ decay, it is possible to measure its half-life and then estimate \mbb.

The sensitivity of a given experiment at the confidence level $n_\sigma$ is expressed by~\cite{delloro}:
\begin{equation}\label{Eq:sensNu}
S^{0\nu} = \frac{\ln 2 ~ N_A ~ \epsilon ~ f_{ab}}{m_A}~ \frac{1}{n_\sigma} ~ \sqrt{\frac{M ~ T}{BI ~ \Delta E}}~.
\end{equation}
This formula emphasizes the role of the experimental parameters needed in the search for the decay: the detection efficiency $\epsilon$, the isotopic abundance $f_{ab}$ of the \bb\ emitter, the target mass $M$, the experimental live-time $T$, the background index $BI$ and the energy resolution $\Delta E$. As shown by the dependence on $\Delta E$ in Eq.~(\ref{Eq:sensNu}), the ability to distinguish the potential signal peak from continuous background is highly dependent on achieving the best possible energy resolution. Furthermore, a precise and stable energy scale is necessary to correctly identify the region of interest around the \qbb\ value and to accurately model the background in that region.
Of particular interest is the case in which $BI$ is sufficiently low that the expected number of background events is less than one count within the region of interest, defined as an energy window centered at \qbb\ and spanning one full-width at half-maximum (FWHM), and a given exposure: this is called ``background-free'' condition.
The \Gerda\ experiment was the first to achieve this condition \cite{gerda_final}, and the design of future next-generation experiments incorporates this goal from the outset. The advantage of this condition is that the sensitivity grows linearly with the experimental mass and time, instead of by square root like in Eq.~(\ref{Eq:sensNu}).

A broad experimental program has been underway for decades. Up to now there has been no observation and the most sensitive results on \onbb\ decay report limits on the half-life up to 10$^{26}$~yr and corresponding upper limits on \mbb\ on the order of 0.1~eV~\cite{legend_2025,kamland2025,cuore2025,gerda_final,mjd_final,EXO-200:2019rkq}.

The Large Enriched Germanium Experiment for Neutrinoless \bb\ Decay (\LEG)~\cite{legend_pcdr} is one of the major current efforts. \LEG\ aims to achieve unprecedented sensitivity by utilizing a modular array of High-Purity Germanium (HPGe) detectors enriched in the candidate isotope \gesix.

To achieve its scientific goals, \LEG\ requires a high degree of precision in its energy reconstruction. This includes achieving an exceptional energy resolution, minimizing non-linearity in the response and ensuring the stability and uniformity of the energy scale across all detectors and over the experiment lifetime.

In this paper, methods related to the energy scale determination of HPGe detectors for the first \Ltwo\ data release on \onbb\ decay search~\cite{legend_2025} are reported. The paper is organized as follows: Section~\ref{Sec:l200} describes the \Ltwo\ detector system and data acquisition organization, Section~\ref{Sec:reconstruction} details the energy estimation process, in Section~\ref{Sec:calibration} the calibration procedure is described, Section~\ref{Sec:performance} reports about extraction of the input parameters needed for the \onbb\ analysis related to the energy scale.


\section{The LEGEND-200 Experiment\label{Sec:l200}}

The \LEG\ collaboration is pursuing an experimental program~\cite{legend_pcdr} with discovery potential at a half-life beyond $10^{28}$~yr using the isotope \gesix.
The first phase, \Ltwo\, is currently ongoing in the former \Gerda~\cite{gerda_hardware,gerda_upgrade} cryostat and water tank at the Gran Sasso Laboratory of the Italian Istituto Nazionale di Fisica Nucleare, located deep underground at 3500~m water-equivalent. \Ltwo\ has a background goal of $2\times 10^{-4}$~\ctsper\ to reach a sensitivity greater than $10^{27}$~yr with 1~\tyr\ of exposure.
The second stage, \Lthou, will operate in a new infrastructure with about 1000~kg of detectors and a background goal of $1\times 10^{-5}$~\ctsper\ to reach a sensitivity greater than $10^{28}$~yr with 10~\tyr\ of exposure. 

\subsection{Experimental Setup}

The \Ltwo\ experiment utilizes HPGe detectors, isotopically enriched in \gesix\ to 86--92\%. These detectors are organized into vertical strings and mounted in low-mass holders consisting of a scintillating PEN plate~\cite{Manzanillas_2022} and underground electroformed copper rods \cite{HOPPE2014116}.
The entire detector array is immersed in a 64~m$^3$ volume of purified liquid argon (LAr). This LAr volume is instrumented to detect scintillation light using wavelength-shifting fibers, which are arranged as two cylindrical curtains surrounding the detector array. Light signals are read out by silicon photomultiplier (SiPM) arrays~\cite{Costa_2023}.

For the first \Ltwo\ data-taking campaign four types of HPGe detectors, for a total of 142~kg, have been deployed. The Inverted Coaxial (\ic) design~\cite{gerda_icpc} is the baseline for \LEG, and it constituted the majority of the mass of the \Ltwo\ array with 86.7~kg. This was complemented by 22.1~kg of P-type Point-Contact (\ppc) detectors from the \MJD\ experiment~\cite{mjd_final}, and 33.7~kg from \Gerda~\cite{gerda_final}, consisting of 14.7~kg of semi-coaxial (\coax) detectors and 19~kg of Broad Energy Germanium (\bege) detectors~\cite{gerda_bege}.
Mirion~\cite{mirion} produced all \bege\ detectors and 73.4~kg of the \ic\ detectors, while the remaining 13.3~kg of \ic\ detectors and all \ppc\ detectors were manufactured by ORTEC~\cite{ortec}.

\subsection{HPGe Detector Readout}

Signals produced by the HPGe detectors are read out by a custom-designed charge-sensitive preamplifier operated in LAr~\cite{Willers_2020}. This is designed to achieve excellent performance: low electronic noise ($<1$~keV FWHM), a fast rise time ($< 100$~ns), high linearity, and a wide dynamic range up to 10~MeV.

The preamplifier system is divided into two stages to meet radio-purity constraints and optimize overall performance. The first stage, the Low Mass Front End (LMFE), is located in the vicinity of the HPGe detector. It is a modified version of the successful \MJD\ solution~\cite{mjd_electronics}.
The LMFE is fabricated on a Suprasil substrate utilizing titanium-gold traces. It hosts the critical input components: a bare-die JFET (Moxtek MX11), a feedback resistor ($R_F \approx 1-2$~G\unit{\ohm}) implemented in amorphous germanium, the feedback capacitance ($C_F \approx 0.4~\text{pF}$) and test-pulse injection capacitance ($C_P \approx 0.1~\text{pF}$) which are both formed by the stray capacitance between the traces.

The second stage of the \Ltwo\ preamplifier is positioned 30--150~cm above the HPGe detectors, with the specific distance for each channel determined by the vertical position of the detector within the string relative to the common support plate at the top of the array.
The design is based on the amplifier employed in \Gerda~\cite{riboldi_redaout} and is fabricated using commercial components on Pyralux circuit boards. This stage employs high-speed, low-noise operational amplifiers to convert the single-ended signal into a differential output, providing enhanced immunity to electromagnetic interference over the transmission lines. The first and second stages are interconnected by four low-mass, custom-made coaxial cables utilizing specialized copper wires and dielectric materials.

\subsection{Data Acquisition}

The Data Acquisition (DAQ) system employs a Flash Analog-to-Digital Converter (FADC) based on the FlashCam readout system~\cite{Gadola_2015}, which digitizes the detector signals with 16-bit precision at a 62.5~MHz sampling rate.
An event is defined by the digitized signal crossing a threshold, set to an energy equivalent to approximately 20~keV. Each resulting trace consists of 8192 samples (131~\unit{\us}), capturing both the pulse and a pre-trigger baseline interval. DAQ operation is managed by ORCA (Object-oriented Real-time Control and Acquisition)~\cite{orca_ref}.

To optimize data transfer and subsequent data processing, the raw digitized traces are divided into two distinct traces during the conversion from ORCA to LGDO (\LEG\ Data Objects) format~\cite{lgdo,daq2lh5}, a custom Python data format developed by the \LEG\ collaboration. For applications that do not require the full sampling resolution, such as energy reconstruction, the traces are down-sampled. This involves summing 8 consecutive samples. The resulting waveforms consist of 1024 samples with a bin width of 128~ns. This significantly reduces the data volume while maintaining sufficient information for precise energy measurement.
For analyses requiring high timing resolution, such as pulse shape discrimination, a second version of the trace is generated at the full sampling frequency, restricted to a window of 1400 samples ($\approx 22$~\unit{\us}) around the signal rise.
This trace contains the signal leading edge to preserve the critical timing information.
These traces from all detectors are stored on disk for subsequent offline analysis.

\begin{figure}
 \centering
    \includegraphics[width=0.5\textwidth]{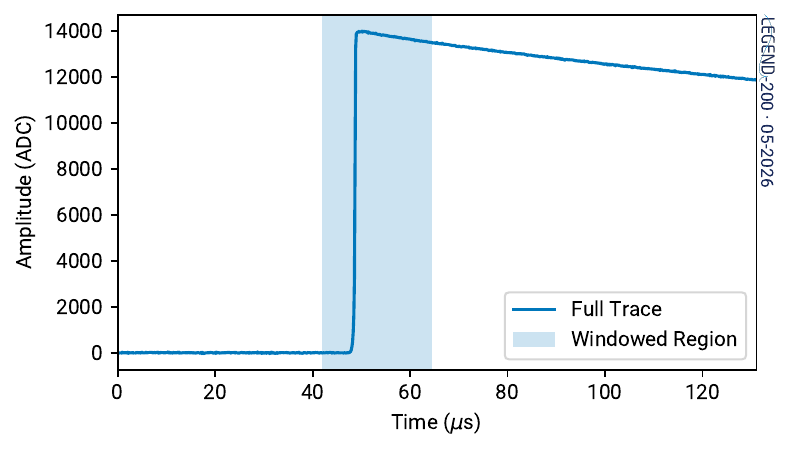}
    \caption{Typical detector waveform acquired in \Ltwo\ corresponding to an energy of approximately 2~MeV. The signal exhibits a fast rise time and a subsequent slow exponential decay.}
    \label{Fig:waveforms}
\end{figure}

Fig.~\ref{Fig:waveforms} illustrates the typical waveform with indication of the windowed region. The digitization captures a baseline region $\approx 48~$\unit{\us} long before the arrival of the detector signal. Following this baseline, the charge signal exhibits a rise time of approximately 1--2~\unit{\us}. The signal is then followed by a window of $\approx 80$~\unit{\mu s}, which captures a portion of the slow exponential decay with a time constant ranging from 400 to 600~\unit{\us}, due to the feedback RC circuit.

\subsection{Data Taking and Partitioning}

The energy calibration and the energy resolution estimation of each detector are performed by irradiating with $^{228}$Th sources, with low neutron emission. The sources are inserted in the LAr cryostat through four calibration systems, each moving four $^{228}$Th sources of about 5~kBq activity per source, situated at different heights for homogeneous event distribution~\cite{l200_cal_system}.

A calibration process is performed on a weekly basis, requiring a duration of a few hours. This initiates the subsequent interval of physics data taking. Calibration and subsequent physics data constitute the primary data-taking interval, referred to as a ``run".
A set of contiguous runs is called a ``period", that ends when hardware operations are performed (e.g., adjustment of detector bias voltage, extraordinary maintenance on electronics).

To consistently combine data over an extended period of time while preserving the excellent energy resolution of the HPGe detectors, it is vital to monitor the stability of the energy scale between calibrations and exclude time spans with significant shifts and fluctuations which would contribute to the width of the peaks. 

For the final analysis, the full data acquisition is segmented into stable time-spans for each detector, referred to as ``partitions".
The stability of a partition is primarily assessed by monitoring the test pulse gain, injected into the electronics chain every 20 seconds. This monitoring is used to track the system stability and allows for the exclusion of time intervals with significant gain variations, or to initiate a new calibration run. Data stability is further validated by monitoring the evolution of energy resolution and $\gamma$-line positions during subsequent calibration runs.
Defining these partitions is crucial for determining stable energy reconstruction and calibration parameters, and for the statistical analysis of the \onbb\ decay as described in~\cite{legend_2025}.

The first data-taking campaign of \Ltwo\ ran from March 2023 to February 2024. For the majority of the detectors, the data are segmented into five partitions separated by planned maintenance operations or by special runs, dedicated to the acquisition of characterization data.


\section{Energy Reconstruction}\label{Sec:reconstruction}

In \Ltwo, data processing and offline analysis are implemented within the \pygama\ framework~\cite{pygama}. 
The energy reconstruction is performed using digital signal processing techniques, employing three distinct methods based on applying a shaping filter to the digitized waveforms. Each method is followed by a charge trapping correction (CTC).
Parameters for shaping filters and CTC are optimized for each detector as described in following sections.

\subsection{Shaping Filters}

\begin{figure}
 \centering
    \includegraphics[width=0.5\textwidth]{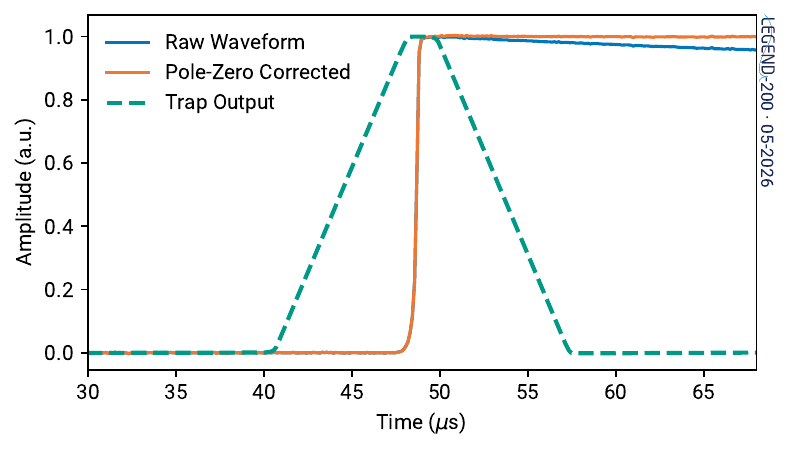}\\
    \includegraphics[width=0.5\textwidth]{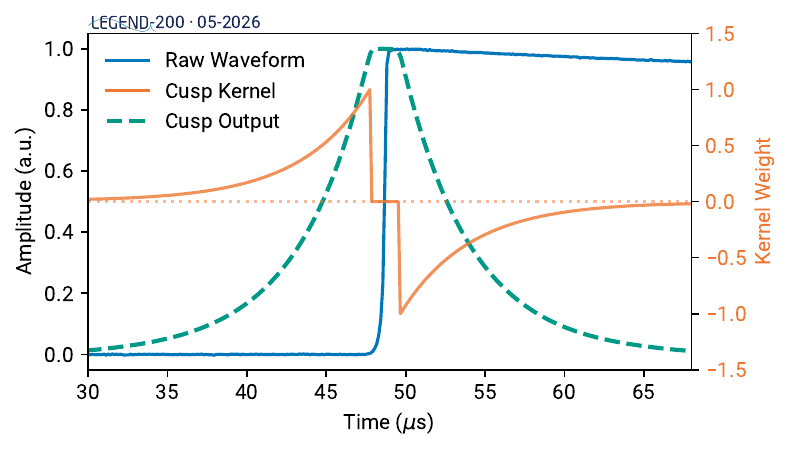}
    \caption{Top: stages of a trapezoidal shaping from the raw waveform to the trapezoidal output. Bottom: example of application of cusp filter. Bottom: illustration of the cusp filtering process for energy reconstruction. The raw waveform (blue) is convolved with the cusp kernel (orange, right axis), producing the cusp output (green dashed line).
    }
    \label{Fig:filters}
\end{figure}

A first energy filtering employs a symmetric trapezoidal filter, a well-established and effective technique for pulse shaping in $\gamma$ radiation spectroscopy~\cite{RADEKA1972525}. This filter is characterized by rising and falling ramps of equal duration, which define the shaping time, and a flat-top in the middle. Its operation can be interpreted as a double integration: the rising and falling ramps act as a finite-time band-pass filter, with their slope determining the degree of high-frequency noise suppression.
The flat-top is introduced to account for the ballistic deficit. It provides a time interval over which the signal amplitude can be accurately determined, as it integrates the non-instantaneous charge collection from the detector and offers robustness against small uncertainties in pulse timing.
Prior to trapezoidal filtering, a pole-zero correction is applied to compensate for the exponential decay of the preamplifier signal. An example of the resulting trapezoidal shaping is shown in Fig.~\ref{Fig:filters} (top).

A technique for enhanced energy resolution was developed for the \Gerda~experiment~\cite{gerda_zac}. It relies on a cusp-like filter featuring a flat-top in the central region and $\sinh$-shaped tails on both sides. The parameter that defines the time scale of this filter is the cusp shaping time. The cusp filter is the optimal shaping filter in the presence of both series and parallel noise~\cite{Deighton1969}. When low-frequency noise and baseline fluctuations are also significant, the energy resolution can be further improved using filters that satisfy the zero-area condition~\cite{GATTI1996117}. This approach was adopted in \Gerda, where energy reconstruction has been performed with the Zero Area Cusp (ZAC) filter, obtained by subtracting two parabolic functions from the sides of the cusp to enforce the zero-area condition.

In \Ltwo, both the standard cusp and the ZAC filters are implemented in the data processing chain. Prior to signal shaping, the filters are convolved with the inverse preamplifier response function, an exponential with a time constant equal to the pole-zero constant ($\tau_{\text{pz}}$), allowing them to be applied directly to the raw signal traces. An example of the resulting filter for the standard cusp, together with the raw waveform and the filter output is shown in Fig.~\ref{Fig:filters} (bottom).

After the application of the shaping filter, the energy of each event is determined as the maximum amplitude of the resulting shaped signal. This approach guarantees that every event receives a valid energy estimate, though it introduces a small systematic bias in the estimator. An optimization procedure, performed separately for each detector (see Section~\ref{Sec:optimization}), is applied to the three available energy estimators. For all detectors the cusp filter was used as the reference method for the main \Ltwo~analysis due to it yielding the best average energy resolution. 

Further developments are ongoing to implement more advanced digital signal processing techniques aimed at achieving optimal shaping under the actual experimental noise conditions and improving energy resolution beyond the current methods~\cite{dplms2023}.

\subsection{Charge Trapping Correction Method}

\ic, \bege\ and \ppc\ detectors deployed in \Ltwo\ have a weighting potential (describing the electrostatic coupling to the electrode~\cite{Ramo_1939}) that is sharply peaked near the p$^+$ contact.
Since electron-hole pairs are generated at a certain distance from the p$^+$ contact, the signal shapes depend on charge carrier drift paths. Events with longer drift have an increased probability to encounter traps from impurities or dislocations in the crystal lattice.
If the trapped charge is not released within the charge collection time, the reconstructed energy is reduced. This leads to a low-energy tail in the peak shape and a degradation of the energy resolution, relevant in particular for the large \ic\ detectors.

Following the procedure described in~\cite{MARTIN201298, mjd_ctc}, energies can be corrected to mitigate degradation from detector resolution by accurately extracting signal drift times. The start time $t_0$ of each pulse is determined by first convolving the pole-zero corrected signal with an asymmetric trapezoidal filter featuring a 128~ns linearly rising section and a 2~\unit{\us} fall time. The $t_0$ is then identified by stepping backwards from the maximum of the filtered signal to the first point that crosses a threshold level set to the baseline Root Mean Square (RMS).
Since this method relies on threshold crossing, a bias toward later times is expected. This filtering preserves the timing information on the rising edge, while reducing the impact of both low and high frequency noise~\cite{mjd_ctc}.


The charge drift parameter $Q_{\text{drift}}$ is estimated by applying a trapezoidal filter, consisting of 4~\unit{\us} rise and fall times and a short 96~ns flat-top. This configuration integrates over the full charge collection period while maintaining sensitivity to variations in the signal leading-edge shape, which reflects the carrier drift dynamics within the crystal. The output is sampled at a fixed time, corresponding to $t_0$ plus the filter length, to ensure complete integration of the charge signal.
The method integrates the area above the leading edge of the signal, which constitutes $Q_{\text{drift}}$. This value provides a parameter for the trapping correction that is perfectly adapted to any pulse shape, independent of the number of interactions inside the detector.

To correct for the described charge trapping effects, we define an effective drift time $dt_{\text{eff}}$. This quantity is calculated as the ratio between $Q_{\text{drift}}$ and the uncalibrated energy of the event. The CTC is then applied using the linear model:

\begin{equation}
    E_{\text{CTC}} = E \left( 1 + \alpha \times dt_{\text{eff}} \right)
\label{Eq:ctc_corr}
\end{equation}
where $E_{\text{CTC}}$ and $E$ are the corrected and uncorrected energies, respectively. The parameter $\alpha$, which indicates the severity of charge trapping, is determined empirically with the optimization procedure described in the following section for each individual detector. The correction is restricted to events above 50~keV. At lower energies, the $t_0$ algorithm is susceptible to baseline noise, causing an unstable drift time estimate that can lead to improper sampling and nonphysical results.

\subsection{Energy Resolution Optimization}
\label{Sec:optimization}

The optimization of energy reconstruction parameters is conducted on a per-detector basis following a two-step procedure. Parameters are first optimized independently for each calibration run acquired during the data-taking period. Following the identification of stable partitions, a common set of optimal parameters are extracted for each detector, then applied to the entire partition for the calculation of the final energy variables.


\subsubsection{Single Calibration Optimization}

The first parameter extracted in the signal processing chain is the pole-zero correction time constant. To determine the appropriate value, a linear fit is performed to the logarithm of the exponential decay tail of each signal. The final constant used for the correction is established by identifying the mode of the distribution generated from these fit results.

The flat-top duration for all shaping filters ($T_{\text{ft}}$) is defined using the time interval required for 99\% of the signal rise, $t_{\text{99}} - t_{\text{0}}$ (where $t_{\text{99}}$ is the time when the amplitude reaches 99\% of the maximum). The specific value of the $T_{\text{ft}}$ is determined from the 99th percentile of the rise-time distribution for events within the 2614.5~keV Full Energy Peak (FEP). 
This percentile-based approach is used to mitigate the influence of noise and outliers, ensuring an optimal flat-top duration. This effectively prevents ballistic deficit while simultaneously maximizing the signal-to-noise ratio.

Subsequently, the shaping filter time constants, generically referred to as $\tau_{\text{filter}}$ (i.e., the ramp duration for the trapezoidal filter, and $\sinh$ parameters for cusp and ZAC filters) are simultaneously optimized together with the CTC parameter ($\alpha$) by performing a systematic grid search and a Bayesian optimization. The goal of this process is to select the parameter combination that maximizes the energy resolution, quantified on the 2614.5~keV FEP.

\begin{figure}
    \centering
    \includegraphics[width=0.5\textwidth]{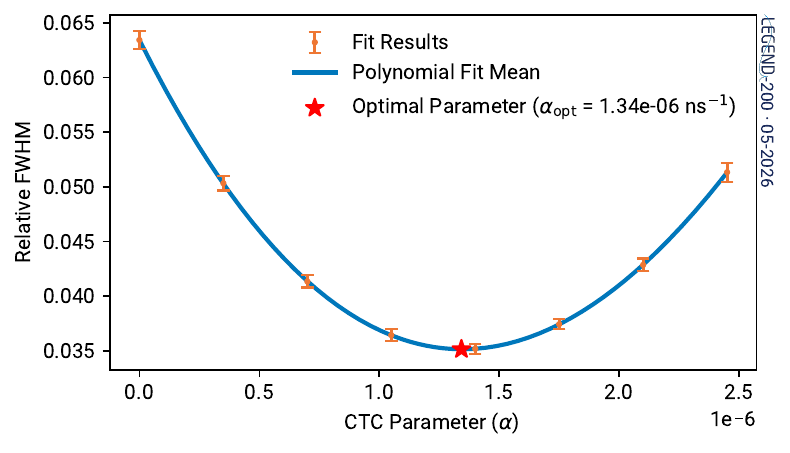}\\
    \includegraphics[width=0.5\textwidth]{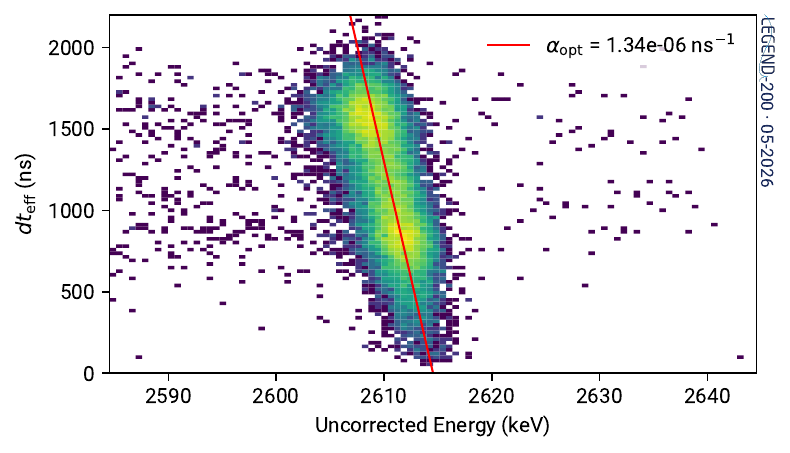}
    \caption{Top: example of CTC parameter optimization. Bottom: effective drift time as a function of the uncorrected FEP energy.}
    \label{Fig:ctc_opt}
\end{figure}
The joint optimization of $\tau_{\text{filter}}$ and $\alpha$ employs a multi-step procedure. An initial coarse scan is performed over a fixed set of shaping time constants: (1, 4, 7, 10, 13, 16)~\unit{\us}. For each shaping parameter, the optimal CTC coefficient ($\alpha_{\text{opt}}$) is determined.
This is achieved by fitting the FEP events using the procedure described in \ref{Sec:peakshape} and calculating the relative FWHM, defined as the FWHM divided by the maximum of the distribution. The relative FWHM is calculated for a range of $\alpha$ from 0 to $3.5 \times 10^{-6}$~\unit{\per\ns}. This metric was chosen to minimize the FWHM without moving events out of the peak. $\alpha_{\text{opt}}$ is determined by fitting a fourth degree polynomial to the relative FWHM curve.

This optimization process is illustrated in Fig.~\ref{Fig:ctc_opt}: the top panel shows the extraction of $\alpha_{\text{opt}}$; the bottom panel shows the effective drift time as a function of the uncorrected FEP energy, illustrating the linear dependence effectively corrected using $\alpha_{\text{opt}}$ in Eq.~(\ref{Eq:ctc_corr}).

Following the energy correction, the FEP is fitted and the Full Width at 20\% Maximum (FW20M) is calculated for each tested parameter set. Unlike the standard FWHM, the FW20M is a more robust metric for characterizing the low-energy tail and overall asymmetry of the peak, that can be symptomatic of residual charge trapping effects or incomplete charge collection.

\begin{figure}
    \centering
    \includegraphics[width=0.5\textwidth]{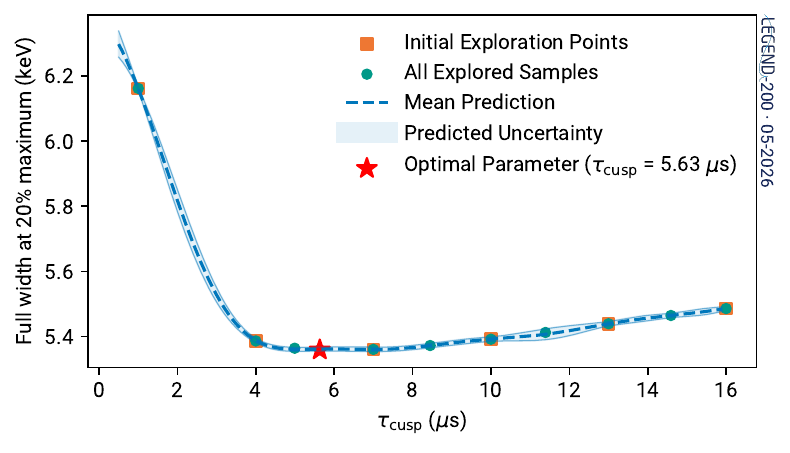}
    \caption{Example of shaping optimization for cusp filter by using the Bayesian optimization algorithm.}
    \label{Fig:filter_opt}
\end{figure}

To determine the global optimal values for $\tau_{\text{filter}}$ and the CTC parameters, a Bayesian optimization algorithm is employed. The FW20M serves as the metric within this optimization.
The algorithm uses a Gaussian process regressor to model the relationship between the filter parameters and the resulting FW20M. After an initial coarse scan of the parameter space, the optimizer iteratively selects the most promising configurations by balancing the refinement of known minima with the exploration of the parameter space.
An example of optimization result for the cusp shaping is shown in Fig.~\ref{Fig:filter_opt}. 

\subsubsection{Partition Parameters}
A common set of filter parameters is determined for the entire partition by calculating the median of the results obtained from the single calibration runs within that partition for each parameter. This averaging step is necessary because parameters extracted from a single calibration run often exhibit lower statistical precision, which results in slightly different values across runs.

Furthermore, if the optimization procedure yields a flat likelihood distribution or a broad minimum, the single-run parameter values can vary significantly. Using a parameter set derived from a single run in the final analysis introduces the risk of systematic instability or bias. By employing the median, which is less susceptible to outliers than the mean, we mitigate the statistical fluctuations and achieve a more stable and statistically representative set of parameters. This final, unified set of parameters is subsequently applied to the full data partition for the calculation of the final energy.

A summary of the key statistical metrics, i.e., mean value, standard deviation and range, is presented in Tab.~\ref{Tab:stats} for each detector type. These metrics are provided for four optimized parameters: pole-zero constant $\tau_{\text{pz}}$, flat-top duration $T_{\text{ft}}$, cusp filter shaping parameter $\tau_{\text{cusp}}$ and charge trapping correction $\alpha_{\text{opt}}$.

\begin{table}
\caption{Summary of energy reconstruction parameters determined at partition levels and combining different detector type.}
\label{Tab:stats}
    \begin{tabular}{c c c c c c}
         & & \coax\ & \bege\ & \ppc\ & \ic\ \\
        \hline
        & Mean & 509 & 475 & 526 & 491 \\
        $\tau_{\text{pz}}$ & $\sigma$ & 40 & 35 & 77 & 39 \\
        (\unit{\us}) & Min & 444 & 434 & 438 & 414 \\
        & Max & 561 & 542 & 801 & 589 \\
        \hline
        & Mean & 1.4 & 1.3 & 1.4 & 2.1 \\
        $T_{\text{ft}}$& $\sigma$ & 0.3 & 0.2 & 0.3 & 0.3 \\
        (\unit{\us}) & Min & 1 & 1 & 1 & 1.4 \\
        & Max & 1.7 & 1.6 & 2.7 & 2.9 \\
        \hline
        & Mean & 11.2 & 5.3 & 4.8 & 5.2 \\
        $\tau_{\text{cusp}}$& $\sigma$ & 3.6 & 1.9 & 2.0 & 2.8 \\
        (\unit{\us}) & Min & 5.2 & 1.3 & 2.2 & 1.0 \\
        & Max & 16.0 & 10.0 & 10.0 & 16.0 \\
        \hline
        & Mean & 0 & 7.6$\times 10^{-7}$ & 1.8$\times 10^{-6}$ & 1.2$\times 10^{-6}$ \\
        $\alpha_{\text{opt}}$ & $\sigma$ & 0 & 4.1$\times 10^{-7}$ & 5.8$\times 10^{-7}$ & 7.3$\times 10^{-7}$ \\
        (\unit{\per\ns}) & Min & 0 & 1.9$\times 10^{-7}$ & 2.0$\times 10^{-7}$ & 3.0$\times 10^{-7}$ \\
        & Max & 0 & 2.0$\times 10^{-6}$ & 3.4$\times 10^{-6}$ & 3.2$\times 10^{-6}$ \\
    \end{tabular}
\end{table}

The pole-zero constant $\tau_{\text{pz}}$ exhibits a tight distribution across the four detector types, as it is primarily determined by the physical feedback RC circuit of the preamplifiers. \coax, \bege\ and \ic\ detectors show low standard deviations and similar means (509, 475 and 491~\unit{\us}, respectively). Conversely, the \ppc\ detectors present a higher mean value of 526~\unit{\us} and the largest standard deviation ($\sigma = 77~$\unit{\us}). This suggests a greater non-uniformity in the preamplifier electronic response or contact properties among the \ppc\ channels.

The flat-top parameter $T_{\text{ft}}$ is centered between $1.3$ and $2.1~\mu\text{s}$ for all detector types. \ic\ detectors show the highest mean value of 2.1~\unit{\us}. This is physically expected, as the large dimensions of the \ic\ geometry result in longer drift paths and slower charge collection times. \bege\ detectors exhibit the most uniform and shortest flat-top duration, reflecting their fast charge collection and small drift distances.

The cusp filter parameter $\tau_{\text{cusp}}$ shows a systematic difference for \coax\ detectors, which require longer shaping times with a mean value of 11.2~\unit{\us}. This is related to the fact that Coax detectors have a larger detector capacitance compared to point-contact designs. A larger $\tau_{\text{cusp}}$ acts as a stronger low-pass filter, which is required to reduce the series noise driven by this larger capacitance. BEGe, PPC, and \ic\ detectors cluster at shorter shaping times (4.8 to 5.3~\unit{\us}) due to their small point contacts and correspondingly low capacitance.
However, in a few particular cases where pronounced low-frequency noise is observed, $\tau_{\text{cusp}}$ is found to reach values as low as 1–2~\unit{\us}. This occurs because the filter is optimized to suppress the low-frequency contributions that would degrade resolution at longer integration times.

Finally, the charge trapping correction parameter $\alpha_{\text{opt}}$ is found to be zero for \coax\ detectors, confirming that this correction is not required due to their configuration. The other three types report small non-zero values on the order of $10^{-6}$ and $10^{-7}$~\unit{\per\ns}. \ppc\ detectors exhibit the highest mean value of $1.8 \times 10^{-6}$~\unit{\per\ns}. This matches expectations, as \ppc\ detectors are known to be more susceptible to charge trapping from crystal impurities, thereby requiring a stronger correction to recover the true energy. The \bege\ detectors report the lowest non-zero mean value.


\section{Energy Calibration Procedure}\label{Sec:calibration}

\begin{figure}
 \centering
    \includegraphics[width=0.5\textwidth]{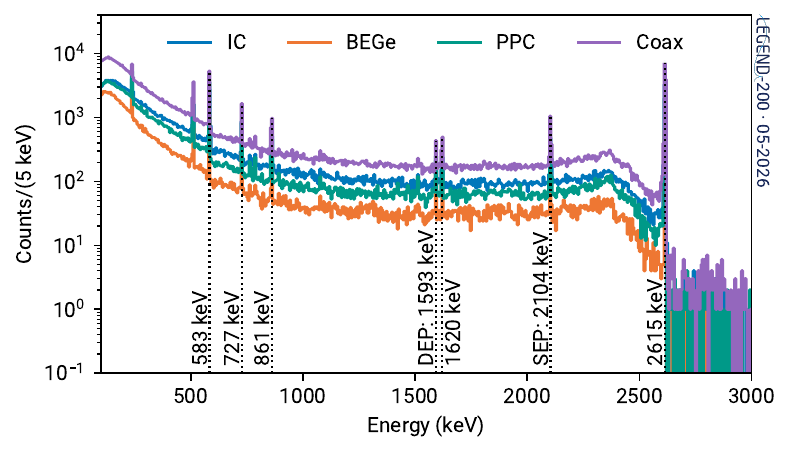}
    \caption{Energy spectra acquired during a weekly calibration for four different germanium detector types: \ic\ (blue), \bege\ (orange), \ppc\ (green) and \coax\ (purple) detectors.}
    \label{Fig:cal_spectra}
\end{figure}

A preliminary energy calibration procedure was initially adapted from the \Gerda\ experiment approach~\cite{gerda_calibration}. The first step involves identifying the prominent $\gamma$-lines from $^{228}\text{Th}$ decay chain within the uncalibrated energy spectrum.
The seven primary lines used for this calibration include the FEPs at 583.2, 727.3, 860.6, 1620.5 and 2614.5~keV, the Double Escape Peak (DEP) at 1592.5~keV and the Single Escape Peak (SEP) at 2103.5~keV.
Fig.~\ref{Fig:cal_spectra} presents an example of calibrated energy spectra acquired during a weekly calibration run for four detectors selected from different detector types. The seven $\gamma$-peaks used for this initial calibration are clearly visible and indicated across all spectra.

It is important to note that the 238.6~keV line from $^{208}\text{Tl}$ and the 511~keV annihilation line are excluded from the fitting procedure. The 238.6~keV line is close to the $241.0~\text{keV}$ line from $^{224}\text{Ra}$, and the 511~keV annihilation peak overlaps with the 510.7~keV line from $^{208}\text{Tl}$, resulting in double-peak structures that complicate precise centroid extraction.

The seven selected $\gamma$-lines located in the spectrum are fitted using the procedure described in detail in \ref{Sec:peakshape}, allowing the precise extraction of the peak centroid $E_{\text{uncal}}$ and its associated uncertainty. Using these centroid positions and their known literature energies, a quadratic calibration curve is derived:
\begin{equation}
    E_{\text{cal}} = a + b E_{\text{uncal}} + c E_{\text{uncal}}^2 ~.
\end{equation}
The inclusion of the quadratic term ($c \ne 0$) is necessary as non-linearities are consistently observed in the detector response.

While the described quadratic procedure works reliably in the majority of cases, failures in parameter estimation have been observed. These failures are often correlated with the relatively low statistics available in the calibration peaks, particularly for smaller detectors. Furthermore, weekly calibration of the full quadratic curve introduces an inherent limitation: statistical fluctuations in the quadratic term $c$ can lead to artificial instabilities in the overall energy scale over time.

\subsection{Two-Stage Calibration Approach}

To mitigate these weaknesses and establish a more robust and stable energy scale, a two-stage energy calibration approach has been developed, following the methodology used in the \MJD~\cite{mjd_calibration}. This approach is based on the assumption that, in the absence of hardware modifications, the primary time dependent change in the system response is the overall gain.

The two-stage procedure separates the time-varying parameters from the stable detector characteristics:
\begin{enumerate}
    \item Stable parameters ($a$ and $c$): the constant offset ($a$) and the quadratic non-linearity term ($c$) are calculated only once for the entire partition using a significantly high-statistics dataset combined from all calibration runs. These terms are assumed to be temporally stable.
    \item Time-varying parameter ($b$): the linear gain term ($b$) is extracted weekly using individual calibration runs. This term effectively accounts for all time-dependent changes in the system gain.
\end{enumerate}
This decoupled approach significantly reduces the statistical uncertainty in the energy scale by fixing the quadratic term and focusing the weekly calibration entirely on the single, dominant time-dependent gain parameter.

\subsection{Gain Calibration}

For the weekly gain correction, a simple and robust method is employed to minimize the probability of mis-calibrations and avoid instabilities. This procedure focuses on accurately determining the linear gain term.

The first step involves obtaining an initial estimate of the energy scale. This is achieved by binning the high-energy portion of the uncalibrated spectrum, specifically from the $95^{\text{th}}$ to the $99.9^{\text{th}}$ percentile of events, and identifying the bin corresponding to the maximum count. This bin is provisionally assigned to the Full Energy Peak (FEP) of the 2614.5~keV line from $^{208}\text{Tl}$, as very few events are expected above this energy in a single calibration run. This initial guess is further refined by binning the whole spectrum and performing a peak search using the seven prominent $\gamma$-lines previously listed. A simple linear fit using these seven positions then provides a more accurate initial gain estimate.

The resulting linear gain term is extracted by performing a peak fit to the 2614.5~keV FEP following the method described in \ref{Sec:peakshape}. The peak position is measured by the mode of the resulting fit function, which is less sensitive to peak tailing than the centroid. The final energy, $E_{\text{cal1}}$, corrected for the weekly gain, is then calculated as:
\begin{equation}
    E_{\text{cal1}} = b_1 E_{\text{uncal}} \label{Eq:cal1}
\end{equation}
where $b_1$ is the calculated gain factor, given by $b_1 = 2614.5~\text{keV} / \text{mode}_{\text{FEP}}$. A full peak fit is performed on the other six calibration lines to extract parameters for auxiliary studies (e.g., energy resolution monitoring), but their positions are not used in the determination of the weekly gain factor $b_1$.

\subsection{Partition Calibration}\label{Sec:part_cal}

The second stage of the calibration is to correct for long-term non-linearities that are assumed to be stable over time. By combining all available calibration data for detectors within the same partition, a high-statistics energy spectrum is generated. This combined spectrum reveals additional, lower-intensity $\gamma$-lines that can be reliably used to extract the non-linear terms.

The intermediate energy $E_{\text{cal1}}$ is then corrected for the partition-specific offset and quadratic non-linearity using:
\begin{equation}
    E_{\text{cal2}} = a_2 + b_2 E_{\text{cal1}} + c_2 E_{\text{cal1}}^2 \label{Eq:cal2}
\end{equation}
where $E_{\text{cal2}}$ is the final calibrated energy used for physics analysis.

To extract the non-linear parameters $a_2$, $b_2$, and $c_2$, fits to all peaks are performed across the high-statistics spectrum. The peak positions are consistently calculated as the mode of the resulting fit function. Peaks are excluded from the fit if the fitting routine does not converge or if the resulting $p$-value is low, typically due to insufficient statistics or overlapping structures. This exclusion criterion usually leaves between 11 and 26 $\gamma$-lines included in the final determination of the partition-level non-linearity.

\subsection{Calibration Parameter Results and Stability}

The described two-stage calibration process has been applied to the entire \Ltwo\ dataset. In the following a description of the resulting calibration parameters ($b_1$, $a_2$, $b_2$, and $c_2$) of Eqs.~\ref{Eq:cal1} and \ref{Eq:cal2} is shown. The method illustrated proved exceptionally robust over the 4000 individual run calibrations performed with zero failures. 


The gain factor $b_1$, which is responsible for converting the uncalibrated energy to keV units, exhibits values around 0.15~keV/ADC. Two \coax\ detectors within the array require a significantly higher conversion factor, with values of $b_1 \approx 0.4$~keV/ADC.
The offset parameter $a_2$ shows values in a range spanning from -1.2 to +1.3~keV, reflecting the necessary low-energy residual correction required to align the energy scale.
The quadratic correction term $c_2$ ranges from $\approx 10^{-8}$ to $10^{-6}$~keV$^{-1}$ and the parameter $b_2$ was found to be tightly constrained to one.

The stability of the energy calibration procedure was initially measured by extracting for each detector the coefficient of variation of the parameters.
The gain parameters $b_1$ and $b_2$ are characterized by a good temporal stability over partitions with variations smaller than 0.1\%. 
The offset parameter $a_2$ and the quadratic term $c_2$ show instead a moderate degree of temporal variation across partitions. For both parameters the relative variation is on the order of 10\%. This instability may be related to the precision on the extraction of the calibration curve when the quadratic correction is very small, leading to statistical wobbling also in the offset term.
Nevertheless, considering the level of these temporal variations and the fact that the \onbb\ decay analysis treats each partition separately, these small instabilities in $a_2$ and $c_2$ are not expected to introduce systematic effects on the final energy scale or energy resolution.

\begin{figure}
 \centering
    \includegraphics[width=0.5\textwidth]{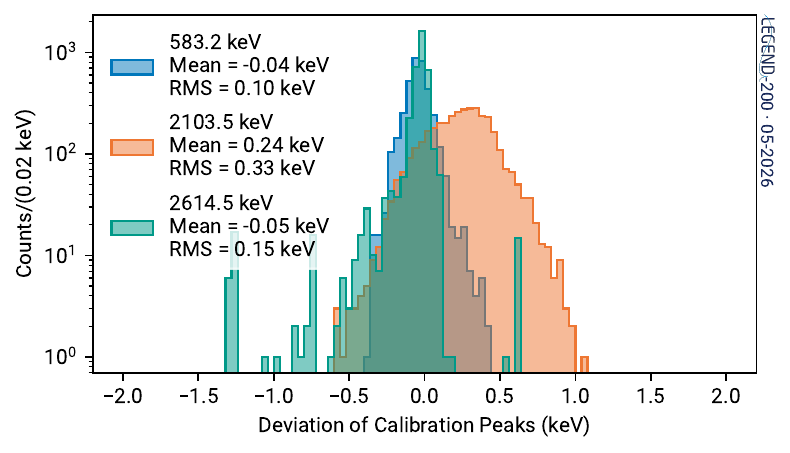}\\
    \includegraphics[width=0.5\textwidth]{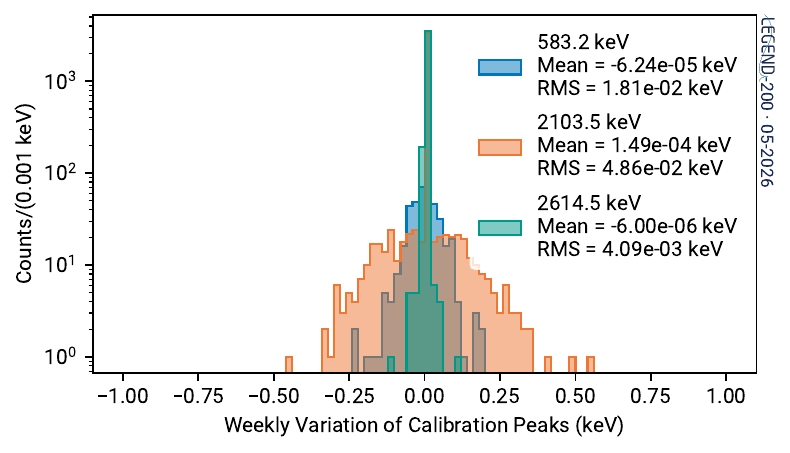}
    \caption{Top: distribution of calibration peak deviations for 583.2, 2103.5 and 2614.5~keV peaks. Bottom: distribution of the energy shifts between subsequent calibration runs for the three peaks.}
    \label{Fig:residuals}
\end{figure}

The overall performance and stability of the derived energy calibration are quantified by analyzing the residuals of the fitted calibration peaks. Fig.~\ref{Fig:residuals} (top panel) reports the distribution of the energy deviation for three prominent $\gamma$-lines (583.2, 2103.5 and 2614.5~keV), calculated across all detectors and all calibration runs. The deviation, $\delta E$, is defined as:
\begin{equation}
    \delta E = E_{\text{true}} - E_{\text{cal}}
    \label{Eq:dev}
\end{equation}
where $E_{\text{true}}$ is the known literature value of the peak energy and $E_{\text{cal}}$ is the calibrated energy extracted from the fit.

The mean and the width (quantified by the RMS) of these distributions are reported in the figure legend. The distributions for both the 583.2~keV and the 2614.5~keV peaks are well-centered near zero (absolute centroid deviation $<0.1~$keV). In contrast, the 2103.5~keV peak distribution exhibits a systematic positive $\delta E$ with a mean value of 0.24~keV and RMS of 0.33~keV. That indicates a non-negligible systematic non-linearity persists in the energy scale around the 2~MeV region, despite the implementation of the quadratic correction. This residual non-linearity may be associated with the integral non-linearity of the FADC. For the \onbb\ decay analysis the energy in the region of interest is corrected taking into account the observed energy deviation which is discussed in Section~\ref{Sec:ebias}.

The bottom panel of Fig.~\ref{Fig:residuals} shows the weekly variation of the calibration peaks, defined as the energy shift between subsequent calibration runs for the same detector channel. The mean values are negligible and the RMS values are very low (e.g., $4 \times 10^{-3}~$keV for 2614.5~keV). This confirms the high temporal stability of the calibrated energy scale, demonstrating that the two-stage calibration effectively mitigates statistical movements between weekly runs.


\section{Energy Scale Performance Metrics}\label{Sec:performance}

The sensitivity of the search for \onbb\ decay hinges directly on the detector performance within the region of interest around \qbb. The two crucial parameters determined by the energy reconstruction and calibration procedures are the FWHM and the energy bias at \qbb. These quantities are essential inputs to the final statistical analysis used to extract the \onbb\ decay half-life limit. Therefore, sustaining these key performance metrics at the demonstrated high standard is fundamental to achieving the physics goals of \LEG.

\subsection{Energy Resolution at \qbb}

The energy resolution at \qbb\ is determined using the combined data from all weekly calibration runs within a specific detector partition.

The resolution at various $\gamma$-lines is initially extracted by fitting the peaks in the combined calibration spectrum, as detailed in Section~\ref{Sec:part_cal}, with the SEP at 2103.5~keV excluded due to Doppler broadening. The FWHM at \qbb\ is then calculated by extrapolating these results using the standard linear model for resolution as a function of energy:
\begin{equation}
    \text{FWHM}(E) = \sqrt{a + b E}~.
\end{equation}

The statistical uncertainty ($\Delta \text{FWHM}_{\text{stat}}$) is quantified using a bootstrap method. This involves generating a set of bootstrapped fit parameters based on the original fit results and their covariance matrix, and then calculating the standard deviation of the resulting FWHM values.

Two primary systematic effects are included in the total uncertainty on the $\text{FWHM}$ at \qbb\:
The first is the temporal resolution stability ($\Delta \text{FWHM}_{\text{shift}}$) that accounts for potential variations in the energy resolution over the data-taking period. It is extracted from the standard error on the mean of the FWHM at \qbb\ determined from the individual, time-separated weekly calibration runs for the partition.
The second systematic uncertainty assesses the dependence on the chosen resolution model function dependence ($\Delta \text{FWHM}_{\text{diff}}$). It is quantified by comparing the FWHM at \qbb\ obtained from the linear fit with the value resulting from an alternative, quadratic resolution model:
\begin{equation}
    \text{FWHM}(E) = \sqrt{a + b E + c E^2}~.
    \label{Eq:eres_quad}
\end{equation}
The introduction of the third parameter, $c$, allows the model to capture energy-dependent non-linearity, such as those arising from incomplete charge collection or specific effects on the detectors.

The total uncertainty on the energy resolution, indicated as $\Delta \text{FWHM}_{\text{TOT}}$, is obtained by summing the statistical and systematic contributions in quadrature:
\begin{multline}
    \Delta\text{FWHM}_{\text{TOT}} = \\
    =\sqrt{\Delta \text{FWHM}_{\text{stat}}^2 + \Delta \text{FWHM}_{\text{shift}}^2 + \Delta \text{FWHM}_{\text{diff}}^2}~.
    \label{Eq:eres_err}
\end{multline}

The majority of BEGe, \ppc\ and \ic\ partitions have uncertainties well below 0.05~keV, reflecting the highly stable and well-calibrated nature of the detectors.

\begin{figure}
 \centering
    \includegraphics[width=0.5\textwidth]{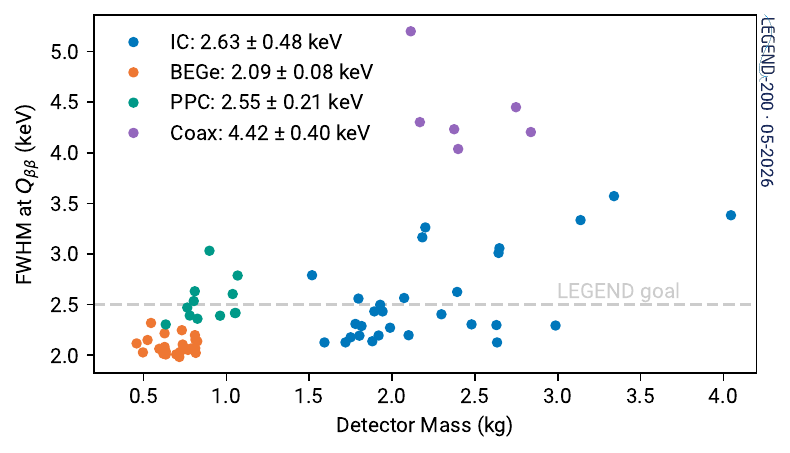}\\
    \includegraphics[width=0.5\textwidth]{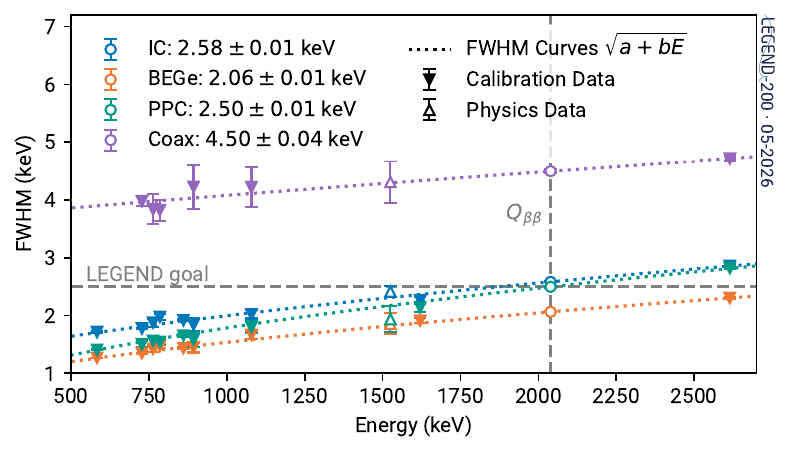}
    \caption{Top: FWHM at \qbb\ as function of the detector mass. For each detector, the reported value is the average over partitions, weighted on partition exposure. The legend shows the average values per detector type, weighted on the exposure of each partition. Bottom: FWHM at \qbb\ per detector type calculated by adding calibration spectra from detectors of that specific type for a selected calibration run. Results are shown together with resolution curves, the FWHM obtained at calibration peaks, and the FWHM obtained in physics data of the $^{42}$K line at 1525~keV.}
    \label{Fig:FWHM_qbb}
\end{figure}

Fig.~\ref{Fig:FWHM_qbb} presents two related perspectives on the energy resolution performance. The top panel shows the FWHM at \qbb\ as a function of the detector mass for each HPGe detector. For a detector composed of multiple partitions, the reported energy resolution is calculated as the average over all its partitions, weighted by the exposure of each partition. The average FWHM values per detector type are reported in the legend.
The \bege\ detectors demonstrate the best FWHM at \qbb\ with average resolution of $(2.09 \pm 0.08)$~keV. \ppc\ and \ic\ detectors also show excellent resolution with an average of $(2.55 \pm 0.21)$~keV and $(2.63 \pm 0.48)$~keV, respectively. A few \ic\ detectors exhibit a larger FWHM, extending beyond 3.0~keV. This degraded resolution has been observed to be related to higher low-frequency noise present in the specific detector string to which these outliers belong. Efforts are currently underway to diagnose and mitigate the source of this higher noise in the affected detectors.
The \coax\ detectors exhibit the worst resolution with an average of $(4.42 \pm 0.40)$ keV. The \LEG\ experimental goal of 2.5~keV at \qbb\ is indicated by a dashed line, showing that the majority of the \bege, \ppc\ and \ic\ detectors meet or exceed this target.
The combined average resolution across all detector types is $(2.47 \pm 0.08)$~keV.


The bottom panel of Fig.~\ref{Fig:FWHM_qbb} reports the energy resolution curves $\text{FWHM}(E)$ obtained by adding the calibration spectra from all detectors of a specific type for a selected calibration run. The dotted lines represent the fit results, while the filled points show the FWHM values extracted from the calibration $\gamma$-lines. The FWHM value extrapolated to \qbb\ is explicitly indicated for each detector type. \bege, \ppc\ and \ic\ detectors show the steepest and best resolution curves.
FWHM values at \qbb\ extracted on resolution curves are different from average values over all partitions reported in the top panel of Fig.~\ref{Fig:FWHM_qbb}, this is expected since they are calculated with different methods. Also plotted for comparison are the FWHM values measured for the $^{42}\text{K}$ $\gamma$-line in physics data for each detector type, demonstrating consistency between the resolutions determined using dedicated calibration sources and those measured using physics background lines.

\subsection{Energy Bias at \qbb}\label{Sec:ebias}

The statistical analysis for the \onbb\ decay search requires a highly accurate energy scale, as the signal is localized close to \qbb. Consequently, a correction for the potential energy bias (or systematic deviation from the true energy scale) is applied to the energy of candidate events, following the established procedure used in \Gerda~\cite{gerda_calibration}.

For each detector partition, the energy bias at \qbb\ and its corresponding uncertainty are evaluated by linearly interpolating the measured energy deviations from the calibration curve at two high-energy $\gamma$-lines: the DEP at 1592.5~keV and the SEP at 2103.5~keV.
This interpolation provides the extrapolated energy bias at the \qbb\ value. The statistical uncertainty on the bias is calculated using a bootstrap method, defined as the standard deviation of a set of extrapolated parameters generated from the fit results and their covariance matrix.

A systematic uncertainty is introduced to account for potential uncorrected energy scale distortions arising from the ADC. Characterization tests reported a $\pm 0.3$~keV non-linearity in the FlashCam digitizer, which is utilized in \Ltwo.
The residual structures observed in both calibration data and dedicated pulser runs are consistent with the magnitude and pattern of the reported ADC non-linearity. Since the standard energy calibration procedure and the subsequent bias calculation do not explicitly model this non-linearity, it must be accounted for as a systematic uncertainty. This term is added in quadrature to the statistical uncertainty to obtain the total energy bias uncertainty.


The resulting energy bias across all detector partitions reports a mean value of $(0.25 \pm 0.01)$~keV. The values are smaller than 1~keV in all cases. This indicates that the global energy scale established by the calibration procedure introduces only a small, quantifiable systematic offset across the array, and the deviation is well-controlled. The total uncertainty on the bias for individual partitions is typically maintained below 0.1~keV. Consequently, for the \onbb\ decay analysis, the energy in the region of interest is corrected by taking into account these observed deviations. This correction ensures that the systematic uncertainty associated with the energy scale is minimized, providing a robust foundation for the final physics results.

To refine the energy scale, dedicated non-linearity studies are planned to fully characterize the electronics response \textit{in situ}. The results of these studies will be used to correct the non-linearity effects in future data releases, which is expected to reduce the magnitude of the measured bias values and further tighten the overall energy scale uncertainty.


\subsection{Comparison to Physics Data}

Physics background data provide an independent cross-check to validate the calibration procedure and the energy resolution model across the detector array. The two most prominent $\gamma$-lines in the background spectrum, originating from $^{40}$K (1460.8~keV) and $^{42}$K (1524.7~keV), are selected for this purpose.

For each detector, the energy resolution is determined by fitting these lines with a Gaussian signal peak superimposed on a linear background. To ensure physical consistency, the background rate is constrained to be non-negative within the fit interval.

\begin{figure}
 \centering
    \includegraphics[width=0.5\textwidth]{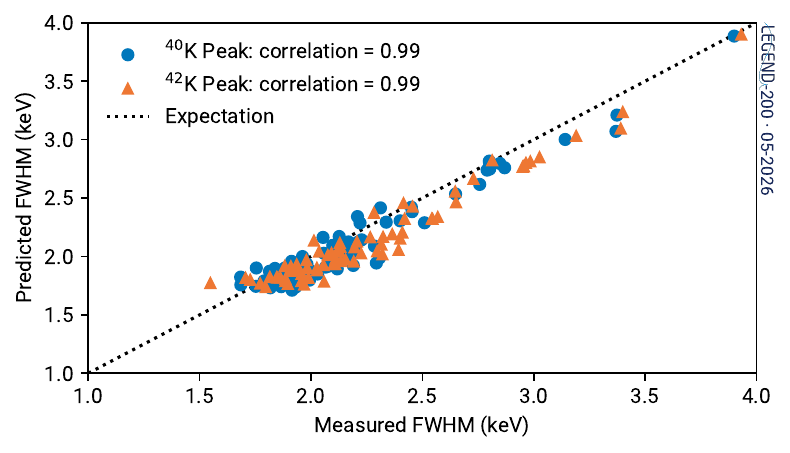}\\
    \includegraphics[width=0.5\textwidth]{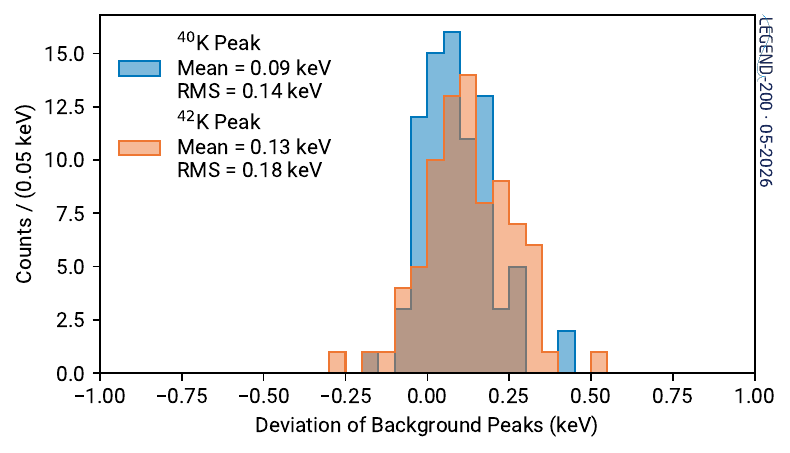}
    \caption{Top: Comparison between the measured energy resolution (FWHM) of the $^{40}$K ($1460.8$ keV) and $^{42}$K ($1524.7$ keV) background lines and the values predicted by the calibration model; each data point represents an individual detector. Bottom: Distribution of the peak deviations for the $^{40}$K and $^{42}$K background lines.}
    \label{Fig:bkg_comparison}
\end{figure}

The measured FWHM values are compared with the resolution derived from calibration runs, presented in the previous sections. As shown in Fig.~\ref{Fig:bkg_comparison} (top), there is a good agreement between the observed K-lines resolution and the predicted values, reporting a Pearson correlation coefficients of 0.99. The measured values are 3--4\% larger than predictions, consistent with long-term instabilities present in physics data-taking that are less pronounced during short calibration runs. This close correspondence confirms that the energy resolution model remains robust and consistent across different data-taking periods.

The peak positions of the two $\gamma$-lines is used to cross-check the energy scale. Fig.~\ref{Fig:bkg_comparison} (bottom) illustrates the distribution of the deviations, as defined in Eq.~(\ref{Eq:dev}), between the fitted peak centroids and their respective literature values. The distributions for both isotopes exhibit a positive bias, with mean deviations of 0.09~keV for $^{42}$K and 0.13~keV for $^{40}$K.
This pattern of energy deviations, consistently observed for both the background lines, provides further confirmation of the systematic non-linearity previously identified at 2103.5~keV in Fig.~\ref{Fig:residuals}.

\section{Summary}

This paper details the energy calibration and performance characterization of the HPGe detectors in the \Ltwo\ experiment, a search for neutrinoless double beta decay. The primary outcome is the validation of the detector system capability to meet the experimental requirements. Key procedures include an advanced energy reconstruction, based on a cusp-like filter and charge trapping correction, to ensure excellent energy resolution and a two-stage calibration for precise energy determination around the \gesix\ $Q$-value (\qbb~$=2039~$keV). The \Ltwo\ detectors demonstrated superior energy resolution, with an average $\text{FWHM}=(2.63 \pm 0.48)$~keV at \qbb\ for \ic\ detectors, the reference detector design for \LEG.
Across the entire detector array, a combined average resolution of $(2.47 \pm 0.08)$~keV at \qbb\ is achieved.
The detectors also exhibited excellent long-term stability, with weekly calibration peak variations remaining below 0.05~keV.
The performance confirms the HPGe detector array is operating at a state-of-the-art level, providing the foundation for \Ltwo\ to significantly advance the search for \onbb\ decay.

\section*{Acknowledgments}
This material is based upon work supported by the U.S. Department of
Energy, Office of Science, Office of Nuclear Physics under Federal Prime 
Agreements DE-AC02-05CH11231, DE-AC05-00OR22725, LANLEM78, and under
award numbers DE-FG02-97ER41020, DE-FG02-97ER41033, DE-FG02-97ER41041,
DE-FG02-97ER41042, DE-SC0012612, DE-SC0014445, DE-SC0017594, DE-SC0018060,
and DE-SC0022339. We acknowledge support from the Nuclear Precision
Measurements program of the Division of Physics of the National Science 
Foundation through grant numbers PHY-1812356, PHY-1812374, PHY-1812409,
PHY-2111140, PHY-2209530, and PHY-2312278, and from the Office of
International Science and Engineering of the National Science Foundation
through grant number OISE-1743790. We gratefully acknowledge the
support of the U.S. Department of Energy through the LANL, ORNL, and
LBNL Laboratory Directed Research and Development (LDRD) Programs. This
research is funded in part by the Deutsche Forschungsgemeinschaft
(DFG, German Research Foundation)—Excellence Cluster ORIGINS EXC
2094-39078331; SFB1258-283604770. We acknowledge the support of the 
German Federal Ministry of Research, Technology and Space (BMFTR)
through grant number 05A2023 and the Max Planck Society (MPG). This work
is supported in part by the European Research Council (ERC) under the
European Union’s Horizon 2020 research and innovation programme (Grant
Agreement number 786430—GemX).  We gratefully acknowledge the financial
support of the Italian Istituto Nazionale di Fisica (INFN); the Polish
National Science Centre (NCN, grant number UMO-2020/37/B/ST2/03905);
the Polish Ministry of Science and Higher Education (MNiSW, grant
numbers DIR/WK/2018/08 and No. 2022/WK/10); the Czech Republic Ministry
of Education, Youth and Sports number LM2023063; the Slovak Research
and Development Agency grant number APVV-21-0377; and the Swiss
National Science Foundation (SNF) numbers FLARE 20FL20\_216572, FLARE
20FL20\_232670, and 200020\_219290. This project has received funding
and support from the European Union’s Horizon 2020 research and
innovation programme under the Marie Skłodowska-Curie grant agreement
number 860881-HIDDeN. This work has been supported by the Science and
Technology Facilities Council (STFC), part of U.K. Research and Innovation
(grant numbers ST/W00058X/1 and ST/T004169/1). We acknowledge the support
of the Natural Sciences and Engineering Research Council of Canada,
funding reference number SAPIN-2017-00023. We acknowledge the support
of the Ministry of Education (MOE) of the Republic of China (Taiwan)
and the National Science and Technology Council (NSTC) of the Republic
of China (Taiwan).  This research used resources provided by National
Energy Research Scientific Computing Center (NERSC), a U.S. Department
of Energy Office of Science User Facility at LBNL, and the Oak Ridge
Leadership Computing Facility at Oak Ridge National Laboratory. We thank
the directors and the staff of the Laboratori Nazionali del Gran Sasso
and our colleagues at the Sanford Underground Research Facility for
their continuous strong support of the LEGEND experiment.

\appendix

\section{Peak Shape Modeling}
\label{Sec:peakshape}

To accurately determine the calibration peak position and precisely quantify the energy resolution of the HPGe detectors, a rigorous peak shape model is employed.
This model is essential for fitting the $\gamma$-lines from the $^{228}$Th calibration source, needed to optimize the parameters of the energy estimators (described in Section~\ref{Sec:reconstruction}), to extract the calibration curves (described in Section~\ref{Sec:calibration}) and to calculate the energy resolution at \qbb\ (described in Section~\ref{Sec:performance}).
The peak fitting is performed with dedicated \pygama\ routines~\cite{pygama}.

\subsection{Gaussian and Low-Energy Tail Models}

The primary component of the peak model is a Gaussian function to describe the ideal response of the detector:

\begin{equation}
g(E) = \frac{n}{\sqrt{2\pi}\sigma} \exp\left(-\frac{(E-\mu)^2}{2\sigma^2}\right)
\label{Eq:gauss}
\end{equation}
where $n$, $\mu$, and $\sigma$ represent the intensity, peak position, and width of the Gaussian, respectively.

Accurate modeling of the peak structure also requires accounting for effects that distort the ideal Gaussian shape. A step function is included to model the continuum background that appears immediately above or below the peaks due to multiple Compton scatters that deposit most, but not all, of their energy:
\begin{equation}
f_{\text{step}} = \frac{a}{2} \text{erfc}\left(\frac{E-\mu}{\sqrt{2}\sigma}\right)
\label{Eq:step}
\end{equation}
where $a$ is the height of the step and $\text{erfc}$ is the complementary error function.

\begin{figure}
 \centering
    \includegraphics[width=0.5\textwidth]{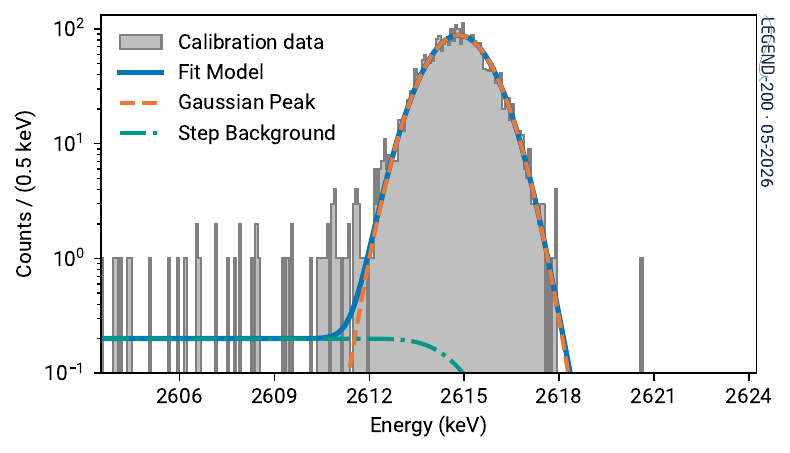}\\
    \includegraphics[width=0.5\textwidth]{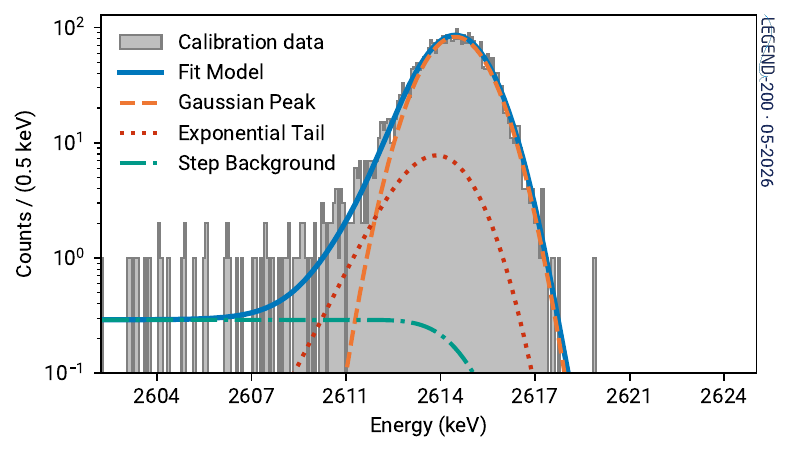}
    \caption{Peak shape fits with Gaussian on step model (top) and including also the low-energy tail (bottom).}
    \label{Fig:peak_fit}
\end{figure}

Finally, a low-energy tail component is incorporated to accurately fit the low-side asymmetry arising from effects such as incomplete charge collection and the residual presence of signal pile-up events. This tail is defined by:
\begin{equation}
f_{\text{tail}} = \frac{b}{2c} \exp\left(\frac{E-\mu}{c} + \frac{\sigma^2}{2c^2}\right) \text{erfc}\left(\frac{E-\mu}{\sqrt{2}\sigma} + \frac{\sigma}{\sqrt{2}c}\right)
\label{Eq:tail}
\end{equation}
where $b$ and $c$ are the height and slope of the tail, respectively. The complete model combines these functions to provide an accurate fit to the calibration peaks, as illustrated in Fig.~\ref{Fig:peak_fit}, a simple Gaussian on a step background (top) with the full model including the low-energy tail (bottom).

\subsection{Model Selection and Fit Strategy}

The spectral analysis relies on the \pygama\ framework, which employs unbinned likelihood fits on the peak energies using the \texttt{iminuit} package~\cite{iminuit}. To ensure that the model complexity accurately matches the data features without overfitting, a staged model selection approach is implemented. This process systematically compares two peak models:

\begin{enumerate}
    \item Simple model: the primary Gaussian function of Eq.~(\ref{Eq:gauss}) combined with the step background component of Eq.~(\ref{Eq:step}).
    \item Complex model: the simple model with the addition of the low-energy tail component of Eq.~(\ref{Eq:tail}) to account for effects like incomplete charge collection.
\end{enumerate}

The model selection decision is based on a dual criterion using the goodness-of-fit and the constraint on the parameters of the complex model. Initially, the goodness-of-fit for the simple model is evaluated using a statistical metric, i.e., $p$-value calculation. The more complex model (including the low-energy tail) is only adopted if the simple model provides an inadequate description of the data (i.e., its $p$-value is below a pre-defined threshold), suggesting that the low-energy asymmetry is a significant feature.

Furthermore, a critical check is performed on the fit results from the complex model. If the uncertainty on the low-energy tail parameter ($b$ or $c$) is too large, indicating that the tail component is not well constrained by the data, the simple model is selected instead. This robust and flexible approach effectively avoids overfitting by only introducing the complex component when both statistically necessary and reliably constrained, ensuring the most appropriate peak shape parameters are consistently determined across all detectors and energy lines.

\bibliography{gen/references}

\end{document}